\documentclass[aps,pre,superscriptaddress,onecolumn,showpacs,floatfix]{revtex4}
\usepackage{epsfig,graphicx,latexsym}

\begin{document}

\title{The effects of  dielectric disorder on van der Waals interactions
in slab geometries}

\author{David S. Dean}
\affiliation{Kavli Institute for Theoretical Physics, University of California Santa Barbara, California 93106, USA}
\affiliation{Laboratoire de Physique Th\'eorique, IRSAMC, Universit\'e Paul Sabatier, 118 Route de Narbonne, 31062 Toulouse Cedex 4, France}

\author{Ron R. Horgan}
\affiliation{Kavli Institute for Theoretical Physics, University of California Santa Barbara, California 93106, USA}
\affiliation{DAMTP, CMS, University of Cambridge, Cambridge,  CB3 0WA, United Kingdom}

\author{Ali Naji}
\affiliation{Kavli Institute for Theoretical Physics, University of California Santa Barbara, California 93106, USA}
\affiliation{Department of Physics, Department of Chemistry and Biochemistry, 
and Materials Research Laboratory, University of California, Santa Barbara, California 93106, USA}
\affiliation{School of Physics, Institute for Research in Fundamental Sciences (IPM), P.O. Box 19395-5531, Tehran, Iran}

\author{Rudolf Podgornik}
\affiliation{Kavli Institute for Theoretical Physics, University of California Santa Barbara, California 93106, USA}
\affiliation{Department of Physics, Faculty of Mathematics and Physics, University of 
Ljubljana and Department of Theoretical Physics, J. Stefan Institute, SI-1000 
Ljubljana, Slovenia}
\affiliation{Laborarory of Physical and Structural Biology, National 
Institutes of Health, Maryland 20892, USA}

\pacs{03.70.+k, 03.50.De, 05.40.-a, 77.22.-d}

\begin{abstract}

We study the thermal Casimir effect between two thick slabs composed of 
plane-parallel layers of random dielectric materials interacting across an 
intervening homogeneous dielectric. It is found that the effective interaction 
at long distances is self averaging and its value  is given by a that 
between non-random media with the effective dielectric tensor of the 
corresponding random media. The behavior at short distances becomes random 
(sample dependent) and is dominated by the local values of the dielectric 
constants proximal to each other across the homogeneous slab. These results
are extended to the regime of intermediate slab separations by using perturbation
theory for weak disorder and also by extensive numerical simulations for a number of 
systems where the dielectric function has a log-normal distribution.

\end{abstract} 

\maketitle

\section{Introduction}

Systems with spatially inhomogeneous dielectric constants exhibit effective van der Waals 
interactions arising from the interaction between fluctuating dipoles in the 
system \cite{mah1976,par2006}. These fluctuation interactions have two distinct 
components: (i) a classical or thermal component due to the zero frequency 
response of the dipoles and (ii) a quantum component due to the non-zero 
frequency/quantum response of the dipoles. Despite the clear physical 
differences in these contributions, the mathematical computation of the 
corresponding interaction is almost identical and boils down to the computation 
of an appropriate functional determinant. The full theory taking into account 
both of these component interactions is the celebrated Lifshitz theory of van 
der Waals - dispersion interactions \cite{dzy1961}, based on boundary conditions imposed on 
the electromagnetic field at the bounding surfaces and the 
fluctuation-dissipation theorem for the electromagnetic potential operators. 
From this formulation one can derive the original Casimir interaction \cite{casimir} by 
taking the limit of zero temperature and ideally polarizable bounding surfaces. 
In this respect the Lifshitz theory is nothing but a proper finite temperature 
and realistic boundary conditions generalization of the Casimir interaction. The 
Lifshitz - van der Waals interactions is thus indeed nothing else but the 
thermal Casimir effect.

\begin{figure}

\epsfxsize=0.3\hsize \epsfbox{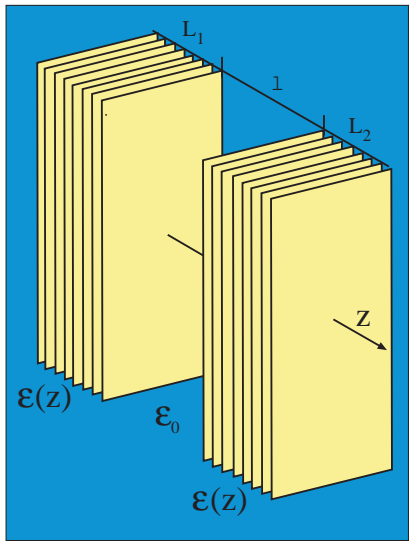}
\caption{(Color online) A schematic presentation of the model. Two finite slabs with disordered 
plane-parallel dielectric layers interacting across a dielectrically homogeneous 
slab of thickness $\ell$. $z$ axis is perpendicular to the plane of the slabs.} 
\label{schematic}

\end{figure}

The major mathematical problems in the computation of Casimir type interactions 
(setting aside the experimental and theoretical challenges to determine the 
correct dielectric behavior) are (i) the application of the Lifshitz approach to 
non-trivial geometries ({\em i.e.} beyond the cases of planar, spherical and cylindrical geometries) and (ii) taking into account local inhomogeneities in the dielectric properties of the media, always present in realistic systems and thus relevant for the comparison of theory with experiment. In this 
paper we will address the second of these points in detail, and to our knowledge we present the 
first analysis of the effect of dielectric disorder on  Lifshitz - van de Waals interactions
(apart from a recent letter  \cite{deanprl} by the authors on this subject). Specifically, we will consider 
the thermal (zero-frequency) Casimir interaction for the case where the local dielectric constant 
is a random variable. Specifically we will consider 
the interaction between two thick parallel dielectric slabs, separated by a 
homogenous dielectric medium, see Fig. (\ref{schematic}). The dielectric response 
within the two slabs is constant in the planes perpendicular to the slab normal, 
but varies in the direction of the surface normal. It is well known that this 
problem can be solved in the case where the dielectric constants of the slabs do 
not vary \cite{par2006} and the result can be tentatively applied to the case of 
fluctuating dielectrics constants via an effective medium theory which consists 
of replacing the fluctuating dielectric constant by an effective (spatially 
homogeneous within each of the slabs) dielectric tensor. Naively one might try the 
approximation

\begin{equation}
\epsilon({\bf x})  \to \langle \epsilon\rangle, 
\end{equation} 
where the angle bracket denotes the spatial or ensemble averaged dielectric 
constant within the slab in question. However the most commonly used 
approximation is that, where the local dielectric tensor is replaced by the 
effective dielectric tensor \cite{mah1976,par2006}, {\em i.e.}

\begin{equation}
\epsilon_{ij}({\bf x}) \to \epsilon^{(e)}_{ij}, \label{eqdiap}
\end{equation}
where the bulk dielectric tensor is defined via 
\begin{equation}
\epsilon^{(e)}_{ij}\langle E_j\rangle = \langle \epsilon_{ij} E_j\rangle.
\end{equation}

The use of the effective dielectric constant is not easily justifiable 
mathematically as an approximation, although physically the effective dielectric 
constant clearly does capture the bulk response to constant electric fields. We 
shall see in this paper, for the random layered dielectric model studied here, 
that the effective dielectric constant approximation of Eq. (\ref{eqdiap}) does 
in fact give the correct value of the thermal Casimir interaction when the two 
slabs are widely separated. On can argue that this  is to be expected on physical grounds as the 
fluctuating electromagnetic field modes with small wave-vector 
(corresponding to variations on large scales) dominate the Casimir interaction for
large inter-slab separation.  The dielectric response of the material to a 
constant electric field is given by the effective dielectric constant and if the 
wave-vector dependent response is suitably analytic near ${\bf k}=0$ we expect that 
$\epsilon^{(e)}_{ij}({\bf k})\sim \epsilon^{(e)}_{ij}(0)=\epsilon^{(e)}_{ij}$ 
for $|{\bf k}|\ll 1$.

\section{\label{model}The model and general analysis}

\subsection{Formulation}

The Hamiltonian associated with the thermal  fluctuations of the 
electrostatic field in a dielectric medium is given by the classical electromagnetic field energy

\begin{equation}
H[\phi]= {1\over 2}\int d{\bf x}\  \epsilon({\bf x}) \left(\nabla \phi({\bf x})\right)^2
\end{equation}
and the corresponding partition function is given by the functional integral 
\begin{equation}
Z = \int d[\phi] \exp(-\beta H[\phi]).
\end{equation}

Differences in dielectric constants lead to the thermal Casimir effect which 
arises from the full treatment of the thermal (zero frequency) van de Waals forces in the 
system. Here we will consider layered systems where the dielectric constant 
$\epsilon$ only depends on the $z$ direction $\epsilon({\bf x}) = \epsilon(z)$. 
If we express the field $\phi$ in terms of its Fourier modes in the plane 
perpendicular to $z$, coordinates denoted by ${\bf x}_\perp$ and which we will 
take to be of area $A$, with wave-vector ${\bf k} = (k_x, k_y)$ then the 
Hamiltonian can be written as

\begin{equation}
H = \sum_{\bf k} H_{\bf k}
\end{equation}
with
\begin{equation}
H_{\bf k} = {1\over 2} \int dz\, \epsilon(z) \left({d{\tilde \phi}(z,{\bf k})\over dz}
{d{\tilde \phi}(z,-{\bf k})\over dz}
+ {\bf k}^2 {\tilde \phi}(z,{\bf k}) {\tilde \phi}(z,-{\bf k})\right).
\end{equation}
A direct consequence of this decomposition of the Hamiltonian is that the partition function 
can be expressed as a sum over the partition function of the individual modes $Z_{\bf k}$ as
\begin{equation}
\ln(Z) = \sum_{\bf k} \ln(Z_{\bf k})
\end{equation}
where
\begin{equation}
Z_{\bf k} = \int d[X] \exp\left(-{1\over 2} \int dz\  \epsilon(z) \left[\left({dX\over dz}\right)^2 
+ k^2 X^2\right]
\right).
\end{equation}
Here $k=|{\bf k}|$ and we have taken into account that the field $\phi$ is real. 

\subsection{Evaluation of the functional integral}
The problem of computing the interaction between slabs composed of layers of 
finite thickness can be studied using a transfer matrix like method 
\cite{pod2004}. However we will use a method based on the Feynman path integral 
instead, which is particularly well suited to the study of systems where the 
dielectric constant can vary continuously in only one direction \cite{pvv}. If we specify the starting 
and finishing points of the above path integral to be $x$ and $y$ respectively at {\em times}
$z'$ and $z$
we see that it has to be of harmonic oscillator form  defined by
\begin{equation}
K(x,y;z',z) = \int_{X(z')=x}^{X(z)=y} d[X] \exp\left(-{1\over 2} \int_0^z dz\ M(z) \left[\left({dX\over dz}\right)^2 + 
\omega^2 X^2\right]
\right),
\end{equation}
where the mass, which is $z$ dependent, is given by $M(z) =\epsilon(z)$ and the 
frequency $\omega$ is given by $\omega = k$. In the case where $M$ and $\omega$ 
are constant, the propagator $K$ is given by the well known formula
\begin{equation}
K(x,y;z',z, M,\omega) 
= \left(M\omega \over 2\pi \sinh(\omega (z-z'))\right)^{1\over 2} 
\exp\left[-{1\over 2}\left((x^2 + y^2)M\omega \coth(\omega (z-z')) Ñ 2xy M\omega {\rm cosech}
(\omega (z-z'))\right)\right]
\end{equation}

In the case where $M$ (or indeed $\omega$) vary with $z$ we can still formally 
compute the path integral via the generalized Pauli - van Vleck formula which 
tells us that $K$ must have the general form
\begin{equation}
K(x,y;z',z) = \left( {b\over 2\pi}\right)^{1\over 2}\exp\left(-{1\over 2} a_i(z',z)x^2 -{1\over 2} a_f(z',z)y^2 +
b(z',z) xy\right).
\label{eq:kernel}
\end{equation}
We may now write down an evolution equation for the coefficients $a_i,\ a_f$ 
and $b$ using the Markov property of the path integral (in fact this is how 
one can prove the generalized Pauli - van Vleck formula \cite{pvv})
\begin{equation}
K(x,y;z',z+\zeta)= \int dw~ K(x,w;z',z) K(w,y;z,z+\zeta).
\end{equation}

Now if we take $\zeta=\Delta z$ infinitesimal and assume that $M$ and $\omega$ are 
constant over $(z, z+\Delta z)$ (but could have jumped at $z$),  then by looking 
at the coefficients of $x^2,\ y^2$ and $xy$ in the so computed $K(x,y;z+\Delta 
z)$ we find the following evolution equations for $a_i,\ a_f$ and $b$ :

\begin{eqnarray}
{\partial a_i(z',z)\over \partial z} &=& -{b^2(z',z)\over M}, \\
{\partial b(z',z)\over \partial z} &=&  -{b(z',z)a_f(z',z)\over M}, \\
{\partial a_f(z',z)\over \partial z} &=& M(z)\omega^2(z) - {a_f^2(z',z)\over M(z)}\;. \label{eqaf}
\end{eqnarray}
Note that the evolution equation for $b$ can also be obtained by examining the 
change in the pre-factor term of the propagator $K$.  An equivalent and related way of
deriving these equations is to note that, from the Feynman - Kac formula,  $K(x,y;z,z')$ satisfies the Euclidean version of 
the Schr\"odinger equation:
\begin{equation}
\frac{\partial}{\partial z}K(x,y;z',z) = 
\left(\frac{1}{2M(z)}\frac{\partial^2}{\partial^2 y} - \frac{1}{2}M(z)\omega^2y^2\right)K(x,y;z',z).
\end{equation}
Substitution for $K(x,y;z',z)$ from Eq. (\ref{eq:kernel}) then yields the  evolution equations (\ref{eqaf}).
As the action is positive definite we expect that both $a_i$ and $a_f$ are 
positive (by considering paths staring or ending at $0$). Also for $|z-z'|$ large, with respect to the correlation length of the disorder, we expect $K$ to factorize in its $x$ and $y$ dependence and thus
the coefficient $b$ should  decay to zero for sufficiently thick slabs.  Note that if $\epsilon$ is a 
stationary process then the path integral should be the same run backwards in 
time, $z$, as when it is run forwards. This means that $a_i$ and $a_f$ should 
have the same steady state distribution.

Consider a system of two thick slabs of respective thicknesses $L_1$ and $L_2$
separated by a distance $l$ and the region between them occupied by a dielectric medium 
of dielectric constant $\epsilon_0$ -- a vacuum or air, for example. From our discussion 
above, for large $L_1$ and $L_2$ the partition function for the mode ${\bf k}$ is thus 
proportional to

\begin{eqnarray}
  Z_{\bf k} &=&\int dx dy \ \exp(-{1\over 2} a^{(1)} _f(k)x^2)  
\left(\epsilon_0 k \over 2\pi \sinh(k z)\right)^{1\over 2} 
\exp\left[-{1\over 2}\left((x^2 + y^2)\epsilon_0 k \coth(k z) - 2xy \epsilon_0 k 
{\rm cosech}(k z)\right)\right] \nonumber \\
&\times& \exp(-{1\over 2} a^{(2)}_ f(k) y^2) \;,
\end{eqnarray}
where $a^{(1)}_f (k)$ is the solution of Eq. (\ref{eqaf}) with some initial conditions (which for 
an infinite slab do not have to be specified)  given at  $z=z'=-L_1$ 
to zero with $\omega=k$ and $M(z) =\epsilon(z)$ and $a^{(2)} _f(k)$ is the 
corresponding quantity for the slab 2 (with initial conditions given at  $z=z'= L_2+l$ and evaluated at  
$z=l$).

We thus find that the $l$-dependent part of the free energy of the mode 
$\bf k$ (up to a bulk term which can be subtracted off to get the effective 
interaction.) is given by
\begin{equation}
F_{\bf k} = {k_B T\over 2} \ln\left( 1- {(a_f^{(1)}(k) - \epsilon_0k) (a_f^{(2)}(k) - \epsilon_0 k) \over
(a_f^{(1)}(k) +\epsilon_0k)(a_f^{(2)}(k) + \epsilon_0k)}\exp(-2kl)\right)\;,
\label{freq}
\end{equation}
with the total $l$ dependent free energy given by
\begin{equation}
F = \sum_{\bf k} F_{\bf k} .
\end{equation}

In order to evaluate the integrals of  $a^{(1,2)}_f(k)$, one first has to solve equations of motion Eqs. (\ref{eqaf}) to get the $z$ dependence of  $a_f(k, z)$ and then proceed to the integrals that enter Eq. (\ref{freq}). The evolution equation for $a_f(k)$ for either slab can be read of from 
Eq. (\ref{eqaf}) and is given by
\begin{equation}
{da_f(k,z)\over dz} =\epsilon(z) k^2 - {a_f^2\over \epsilon(z)},
\label{equndef}
\end{equation}
where we have dropped the explicit dependence on the initial point $z'=-L_1$. 
Though it does not look like it at the first sight, this equation is simply a rewriting of the underlying 
Poisson equation for the original, charge free, dielectric system . This can be seen as follows.
Assume first of all that $a_f(k,z)$ can be parameterized with a function $\psi(k,z)$ as
\begin{equation}
a_f(k,z) = \epsilon(z) {d\over dz}\ln{\psi(k,z)}.
\label{defabs}
\end{equation}
In quantum mechanics problems with disorder the above change of variables is often used since in the presence of disorder nonlinear first order equation is easier to analyze then the second order linear equation \cite{itzykson}. Inserting now this {\em ansatz} back into Eq. (\ref{equndef}) we find that it implies
\begin{equation}
  {d\over dz}\left(\epsilon {d\psi\over dz}\right)  - \epsilon  k^2 \psi = 0,
\end{equation}
for $\psi(k,z)$, which,  is nothing but the  Poisson equation for an inhomogeneous dielectric, where the inhomogeneity is only in the $z$ direction, which has been Fourier
transformed in the directions perpendicular to $z$. The $a_f(k,z)$ in Eq. (\ref{defabs}) is thus given by the solution of the Poisson equation in the specified planar geometry. This is of course no surprise since we are indeed dealing with an inhomogeneous electrostatic problem. On the other hand the derivation presented above is completely equivalent to the transfer matrix method \cite{pod2004} or to the density functional method \cite{veble} for evaluating the van der Waals forces. One of the clear strengths of
this method is that it allows the of computation  the van der Waals interaction to be carried out using
a local method where the coefficients $a_f$, $a_f$ and $b$ for any of the media involved can be
computed and then the interactions between any combinations of media can be worked out 
in terms of  these coefficients.

If we now write $a_f^{(i)}(k,z) = k\alpha^{(i)}(k, z)$ and if the distributions of 
the $\alpha^{(i)}(k, z) = y$ are given by $p_i(k,y)$ then we find that, in three 
dimensions the average $l$ dependent free energy is given by
\begin{equation}
\langle F \rangle = {k_B TA\over 4\pi }\int dk\, k \int dy_1\int dy_2 \, p_1(k,y_1)p_2(k,y_2)
\ln\left( 1- {(y_1 - \epsilon_0) (y_2 - \epsilon_0 ) \over
(y_1 +\epsilon_0)(y_2 + \epsilon_0)}\exp(-2kl)\right),\label{free1}
\end{equation}
where the angled bracket on the l.h.s. indicates the disorder average over the 
dielectric constant within the slabs and we have assumed that the realizations 
of the disorder in the two slabs are independent. This is why the joint disorder 
probability distribution is multiplicative for the two layers.

In the case where the slab thickness is large, $L \to \infty$, it is simple to derive 
a scaling formula from Eq. (\ref{equndef}) for the probability distribution.
Let $\epsilon(z) = \epsilon_g f(z)$ where $f(z)$ is an instance drawn from an 
ensemble distributed according to a given distribution. Then $p(k,y;\epsilon_g)$ 
is the resulting distribution for $y$ given the mode wave-vector $k$. We then 
find that 
\begin{equation}
p(k,y;\lambda\epsilon_g) = \frac{1}{\lambda}p(k,y/\lambda;\epsilon_g) \;.
\label{pscale}
\end{equation}
Thus, we need to compute the probability distribution once only, say for $\epsilon_g=1$,
and obtain those for other values of $\epsilon_g$ using this scaling result.

\subsection{Large $l$ limit}

Let us first investigate the form of van der Waals interaction free energy in 
the limit of large separations between the two slabs. The equation obeyed by 
$\alpha$ is
\begin{equation}
{d\alpha(k,z)\over dz} =\epsilon k - {k\alpha^2\over \epsilon} \label{eqmal},
\end{equation}
which can be written as
\begin{equation}
{d\alpha(k,\zeta)\over d\zeta} =\epsilon({\zeta/ k})  - {\alpha^2\over \epsilon({\zeta/ k})},
\end{equation}
with $\zeta = zk$. When $k$ is small $\epsilon({\zeta/ k})$ varies very 
rapidly and thus becomes de-correlated from the value of $\alpha$.  The Laplace 
transform for the probability density function of $\alpha$ is defined by
\begin{equation}
{\tilde p}(k,s,\zeta) = \int_0^\infty dy \exp(-sy) p(k,y,\zeta) = 
\langle \exp(-s \alpha(k,\zeta))\rangle\;,
\end{equation}
and, from the equation of motion Eq. (\ref{eqmal}), obeys
\begin{equation}
-{1\over s} {d{\tilde p}(k,s,\zeta) \over d\zeta} = -{1\over s} {d\over d\zeta} \langle \exp(-s \alpha(k,\zeta))\rangle\; = 
\left\langle \epsilon({\zeta/ k})  \exp(-s \alpha(k,\zeta))  - 
{\alpha^2\over \epsilon({\zeta/ k})} \exp(-s \alpha(k,\zeta))\right\rangle  .
\end{equation}
Assuming that $k$ is small and thus $ \alpha(k,\zeta)$ and $\epsilon({\zeta/ k})$ 
are de-correlated we can write
\begin{equation}
-{1\over s} {d{\tilde p}(k,s,\zeta) \over d\zeta} = \langle \epsilon\rangle {\tilde p}(k,s,\zeta)
-\left\langle {1/ \epsilon}\right\rangle {d^2\over ds^2}{\tilde p}(k,s,\zeta)\;.
\end{equation}
This equation has  the large $\zeta$ equilibrium solution (justified as we are taking the limit $L_1,\ L_2\to \infty$) 
\begin{equation}
\lim_{\zeta\to \infty}{\tilde p}(k,s,\zeta)= \exp(-\epsilon^* s)\;,
\end{equation}
with 
\begin{equation}
\epsilon^* = \sqrt{\frac{\langle\epsilon\rangle}{\left\langle {1/ \epsilon}\right\rangle}}\;.  \label{eqstar1}
\end{equation}
Inverting the Laplace transform then gives the equilibrium distribution
\begin{equation}
p(y,k)   = \delta(y- \epsilon^*) \label{psk}
\end{equation}
at small $k$.  When $l$ is large the integral in Eq. (\ref{free1}) is dominated 
by the small $k$ behavior and we may use the analysis presented above, 
especially Eq. (\ref{psk}) in Eq. (\ref{free1}), to give the following 
asymptotic form for the interaction free energy
\begin{equation}
\langle F \rangle(l \longrightarrow \infty) \sim {k_B TA\over 16\pi l^2 }
\int u\, du\,  \ln\left( 1- \Delta_1^*\Delta_2^*\exp(-u)\right) \sim -{H^*\over l^2}\;,
\label{free2}
\end{equation}
with
\begin{equation}
\Delta_i^* = {(\epsilon^*_i - \epsilon_0) \over (\epsilon_i^* +\epsilon_0)}\;,
\end{equation}

and where $\epsilon_i^*$ are defined via Eq. (\ref{eqstar1}). The subscript $i$ 
on the angled brackets signifies that we are averaging the dielectric constant 
in the slab $i$.  The term $H^*$ defines an effective disorder-dependent Hamaker 
coefficient.

This result can be obtained via another more physical argument. The random 
layered materials can be replaced by an effective anisotropic medium where the 
dielectric tensor is non isotropic and has the form

\begin{eqnarray}
\epsilon^{(e)}_{zz} &=& \epsilon_{||}\;, \\
\epsilon^{(e)}_{xx}&=&\epsilon^{(e)}_{yy} = \epsilon_{\perp}\;,
\end{eqnarray}
all other terms being zero by symmetry. The term $\epsilon_{||}$ is the 
dielectric constant in the $z$ direction is known exactly and is given by
\begin{equation}
\epsilon^{(e)}_{||} = \frac{1}{ \left\langle {1/ \epsilon}\right\rangle}\;,
\end{equation} 
and the perpendicular component is simply given by
\begin{equation}
\epsilon^{(e)}_{\perp} = \langle \epsilon \rangle\;.
\end{equation}
The forms of $\epsilon^{(e)}_{||}$ and $\epsilon^{(e)}_{\perp}$ follow simply 
from the fact that in the perpendicular direction the dielectric constant is 
obtained by analogy to capacitors in series and in the parallel direction by 
analogy to capacitors in parallel arrangement \cite{podg2}.

It is a straightforward exercise to see that the effective value of $\epsilon^*$ 
for this system coincides with that of Eq. (\ref{eqstar1}) above. This 
correspondence makes physical sense since the high wave length or small $k$ 
fluctuations of the electric field are responsible for the behavior of the 
Casimir interaction at large distances and the effective dielectric response at 
low (but non-zero $k$) must be close to that of the response to a constant field 
{\em i.e.} described in terms of a dielectric constant.  An interesting 
consequence of this result is that for large separations (where $l$ is much 
greater than the correlation length of the dielectric disorder) the thermal 
Casimir interaction free energy is self averaging.

It is instructive to compare the result for the Casimir interaction at large 
separations in the case of a fluctuating dielectric constant and in the case of 
a homogeneous medium, whose dielectric constant is the average of that in the 
fluctuating medium: $\epsilon_h = \langle \epsilon\rangle$ (the subscript $h$ 
signifying that the medium is homogeneous). The homogeneous medium has a Hamaker 
constant
\begin{equation}
H_h = -{k_B T\over 16\pi }\int u\, du \,
\ln\left( 1- \Delta^2_h\exp(-u)\right),
\end{equation}
with
\begin{equation}
\Delta_h =
{\langle \epsilon\rangle  - \epsilon_0 \over
\langle \epsilon\rangle  +\epsilon_0},
\end{equation}
when the medium (1) has the same composition as the medium (2). Jensen's inequality tells us that
\begin{equation}
\left\langle {1/ \epsilon} \right\rangle \geq {1\over \langle \epsilon\rangle}
\end{equation}
since the function $f(x)=1/x$ is convex. Thus
\begin{equation}
\epsilon^* = \sqrt{\frac{\langle \epsilon\rangle }{ \left\langle {1/ 
\epsilon}\right\rangle}}\; \leq \langle \epsilon \rangle \label{ineq}
\end{equation} 
Clearly the effective Hamaker constant is a monotonic function of $\Delta^2$ and 
the interaction is always attractive. The difference in $\Delta^2$ for the two 
systems is

\begin{equation}
\Delta^{*2} - \Delta_h^2 = 
4\frac{\epsilon_0(\epsilon^*-\langle\epsilon\rangle)(\epsilon^*\langle 
\epsilon\rangle -\epsilon_0^2)} {(\langle\epsilon\rangle 
+\epsilon_0)^2(\epsilon^*+\epsilon_0)^2}\;.
\end{equation}

Therefore, using Eq. (\ref{ineq}), we find that, if $\epsilon^*\langle \epsilon 
\rangle >\epsilon^2_0$ then $H_h>H^* $, and the interaction between the two 
homogeneous media is stronger than that between the two fluctuating media. This 
condition can be written as
\begin{equation}
\frac{\langle \epsilon \rangle ^3 }{\left\langle {1/ \epsilon}\right\rangle }> \epsilon_0^4\;
\end{equation}
and is always satisfied if $\langle1 /\epsilon\rangle^{-1} > \epsilon_0$. 
However it is always violated (the interaction between the homogeneous media is 
weaker than that between the random media) if $\langle \epsilon\rangle 
<\epsilon_0$.  We thus see that, depending on the details of the distribution of 
the fluctuating dielectric response in the two slabs and the dielectric response 
of the medium in-between, the effective interaction at large inter-slab 
separations can be stronger or weaker than that for a uniform medium with a 
dielectric constant equal to the average value of that of the media in which it 
fluctuates.

\subsection{Small $l$ limit}

One would imagine that as the distance between the slabs is reduced the 
result will be increasingly dominated by the slab composition at the two 
opposite faces \cite{par2006}. Indeed in the small $l$ limit Eq. (\ref{free1}) is dominated by 
the large $k$ behavior. The asymptotic behavior can be extracted if one assumes the {\em ansatz} 
\begin{equation}
\alpha(z,k) = \sum_{n=0}^\infty {\alpha_n(z)\over k^n}. 
\end{equation}
Substituting this into Eq. (\ref{eqmal}) gives the following chain of equations for $\alpha_n(z)$
\begin{equation}
{1\over k}\sum_{n=0}^\infty {\frac{1}{k^n}\frac{d\alpha_n(z)}{dz}} = 
\epsilon(z) - {1\over \epsilon(z)}\sum_{n,m=0}^\infty
{\alpha_n(z)\alpha_m(z)\over k^{m+n}}\;.
\end{equation}
>From here it is easy to see that to order $O(1)$ the leading asymptotic result of Eq. (\ref{ask}) is given by
\begin{equation}  
\alpha_0(z) = \epsilon(z)\;.
\end{equation}
The equation for the corrections ($n\geq1$) to this asymptotic limit  is
\begin{equation}
{d\alpha_{n-1}(z)\over dz} = -{1\over \epsilon(z)}\sum_{m=0}^n \alpha_m(z) \alpha_{n-m}(z),
\end{equation}
and the next two terms in the asymptotic expansion are given by
\begin{eqnarray}
\alpha_1(z)  &=&-{1\over 2} {d\epsilon(z)\over dz}\;,\\
\alpha_2(z) &=&  {1\over 4}{d^2\epsilon(z)\over dz^2} - 
\frac{1}{8\epsilon(z)}\left({d\epsilon(z)\over dz}\right)^2\;.
\end{eqnarray}
It is straightforward to realize that these terms generate $O(1/l)$ corrections to the asymptotic result
which are subdominant when $l$ is large. Thus to the leading order
\begin{equation}
\alpha(z,k) \approx \alpha_0(z) = \epsilon(z) 
\label{ask}
\end{equation}
and from here it follows straightforwardly that
\begin{equation}
\lim_{k\to \infty} p_i(y,k) = \rho_i(\epsilon)\label{largek}
\end{equation}
where $\rho_{i}$ is the probability density function of $\epsilon(z)$ in medium 
$i$. This result is easily understood from the physical discussion above.  The 
average of the thermal Casimir interaction free energy Eq. (\ref{free1}) in the 
small separation limit is thus given by
\begin{equation}
\langle F\rangle (l \longrightarrow 0) \sim {k_B T\over 16\pi l^2 }\int u\,du\int 
d\epsilon_1 d\epsilon_2 \, \rho_1(\epsilon_1)\rho_2(\epsilon_2) \,\ln\left( 1- \Delta_1\Delta_2\exp(-u)\right),
\label{freesmall}
\end{equation}
with
\begin{equation}
\Delta_i = {\epsilon_i-\epsilon_0\over \epsilon_i-\epsilon_0}.
\end{equation}

The forms of the thermal Casimir interaction free energy are thus given by Eqs. 
(\ref{freesmall}) and (\ref{free2}) in the small and large inter-slab separation 
limits respectively. We have thus obtained the limiting behavior of the thermal 
Casimir effect in the limit of large separation between the slabs, where the 
free energy is given by self-averaging and thus the distributions of 
$\alpha(k,z)$ are strongly peaked, and in the limit of small separation, where 
the free energy is a random variable that has to be averaged over the 
probability density function of the dielectric constant in the media composing 
the two interacting slabs.

\section{Small disorder limit - perturbation theory}\label{secpert}

The analysis presented above is valid for any type of disorder, irrespective of 
its properties. Here we also investigate a different approach that takes into 
account the disorder effects on a perturbative level {\em i.e.} another way of 
approaching this problem analytically is to assume that the disorder is small. 
We assume in this case that the dielectric response can be written with an {\em 
ansatz} of the form

\begin{equation}
\epsilon(z) =\epsilon_g\exp(\lambda \phi(z))\label{epsg}\;,
\end{equation}
where $\lambda$ is a scalar parameterizing the strength of disorder which will be used as 
the expansion parameter for  perturbation theory. When $\lambda =0$ we have the disorder free
homogeneous system which is the starting point for the perturbation expansion (zeroth order). 
If the mean of the field of $\phi$ is zero then $\epsilon_g$ is the geometric mean of the 
dielectric constant. We now assume that $\lambda$ is small and  write
\begin{equation}
\alpha(z,k) = \sum_{n=0}^\infty \lambda^n\alpha_n(z,k) .
\end{equation}
Substituting this into Eq. (\ref{eqmal}) and matching the powers of $\lambda$, 
we obtain the following equations for the first three terms in the perturbation 
expansion when $z$ is large:
\begin{eqnarray}
\alpha_0 &=& \epsilon_g\;, \\
{d\alpha_1\over dz } &=& k(-2\alpha_1 + 2\epsilon_g \phi)\;, \\
{d\alpha_2\over dz } &=& k(-2\alpha_2 -{\alpha_1^2\over \epsilon_g} + 2\alpha_1\phi).
\end{eqnarray}
At large $z$ we can solve the last two equations to obtain
\begin{eqnarray}
\alpha_1 &=& 2k\epsilon_g \int_0^z dz'\ \exp(-2k(z-z'))\phi(z')\;, \\
\alpha_2 &=& 4k^2\epsilon_g\int_0^z dz'  \exp(-2k(z-z'))\phi(z')\int_0^{z'} dz^{\prime\prime} 
\exp(-2k(z'-z^{\prime\prime}))\phi(z^{\prime\prime}) \nonumber \\
&-& 4k^3\epsilon_g\int_0^z dz \exp(-2k(z-z'))\left [\int_0^{z'} dz^{\prime\prime}
\exp(-2k(z'-z^{\prime\prime}))\phi(z^{\prime\prime})\right]^2\;.
\end{eqnarray}

We can now verify some of our previous results at the level of perturbation 
theory. For $k\to \infty$ we have that $2k\exp(-2k(z-z')) \to \delta(z-z')$, and 
this gives

\begin{eqnarray}
\alpha_1 &\to& \epsilon_g \phi(z)\;, \\
\alpha_2 &\to& {\epsilon_g\over 2} \phi^2(z).
\end{eqnarray}
This gives
\begin{equation}
\alpha \to \epsilon_g( 1+ \lambda \phi + {\lambda^2\over 2}\phi^2) \approx \epsilon_g\exp(\lambda \phi)
=\epsilon(z)
\end{equation}
in agreement with the large $k$ result stated previously.  Also in the limit of 
small $k$ we will have that
\begin{equation}
\lim_{k\to 0}2k\int_0^z dz^{\prime} \exp(-2k(z-z')) \phi(z') \to \langle \phi\rangle\;,
\end{equation}
for large $z$ and so we have
\begin{eqnarray}
\alpha_1 &\to& \epsilon_g \langle \phi\rangle\;,  \\
\alpha_2 &\to& {\epsilon_g\over 2} \langle \phi\rangle^2.
\end{eqnarray}
Since $\langle\phi\rangle = 0$, we find 
\begin{equation}
\lim_{k\to 0}\alpha(z,k) = \epsilon_g\;, 
\end{equation}
which is in agreement (to $O(\lambda^2)$) with the general result Eq. (\ref{eqstar1}).  

Now let us consider the case where the field $\phi$ is Gaussian of zero mean with correlation 
function
\begin{equation}
\langle \phi(z)\phi(z')\rangle = \exp(-\omega|z-z'|),\label{oupr}
\end{equation}

where $1/\omega$ is the correlation length of the field. This Gaussian field has 
the correlation function of an Orstein-Uhlenbeck (OU) process and is Markovian. 
The moments relevant to the degree in perturbation theory we are working to 
($O(\lambda^2)$) are

\begin{eqnarray}
\langle \alpha _1\rangle &=& 0\;, \\
\langle \alpha_1^2 \rangle &=& = {2k\epsilon_g^2\over 2k + \omega}\;, \\
\langle \alpha_2\rangle &=& {k\epsilon_g\over 2k + \omega}\;. 
\end{eqnarray}
Using this we may write the corresponding free energy as a random variable
\begin{equation}
 F  = {k_B T\over 16\pi l^2 }\int u\,du \ln\left( 1- \Delta'_1\Delta'_2\exp(-u)\right)\;,
\end{equation}
where
\begin{equation}
\Delta'_i = {\epsilon'_i - \epsilon_0\over \epsilon'_i + \epsilon_0}\;,
\end{equation}
with
\begin{equation}
\epsilon'_i= \epsilon_{g,i}\left(1 + \lambda\sqrt{ u\over u+\omega_i l} \sigma_i+ 
{\lambda^2\over 2} { u\over u+\omega_i l}\right)\;.
\end{equation}
The subscript $i$ refers to the medium (1, or 2) and the $\sigma_i$ are 
independent Gaussian random variables of zero mean and unit variance. Note that to 
order $\lambda^2$ we can replace the term $\alpha_2$ by its mean and have done 
so in the above.  Also we should expand the corresponding expression for the 
free energy to order $\lambda^2$, this will ensure that the resulting average 
and moments will always be finite.

\section{Numerical simulations}

In this section we perform some numerical simulations to verify the general 
asymptotic behavior of the Casimir free energy that we have analyzed in earlier
sections. For large slab separation (large $l$) the results are  summarized by
Eqs. (\ref{psk}) and (\ref{eqstar1}), and in the small separation limit
(small $l$) we expect Eq. (\ref{largek}) to hold. We present simulations to 
show that these equations are indeed correct in the relevant limit. 

\begin{figure}
\epsfxsize=0.5\hsize \epsfbox{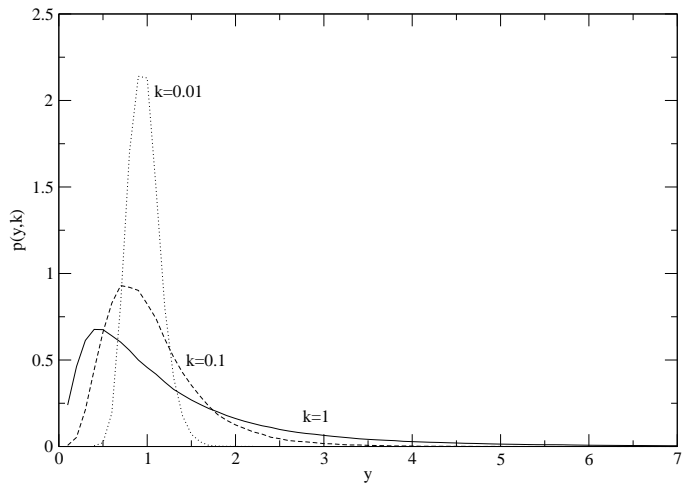}
\caption{Probability density for $\alpha$ for $k=1,\ 0.1, \  0.01$ for $\epsilon(z)=\exp(\phi(z))$ where
$\phi$ is the  OU Gaussian process with correlation function given by Eq. (\ref{oupr}).
The distribution $p(y,k)$ becomes sharply peaked around $y=1$ which is as predicted for small $k$,
$\epsilon^*$ given by Eq. (\ref{eqstar1})  }
\label{oumultik}
\end{figure}

\begin{figure}
\epsfxsize=0.5\hsize \epsfbox{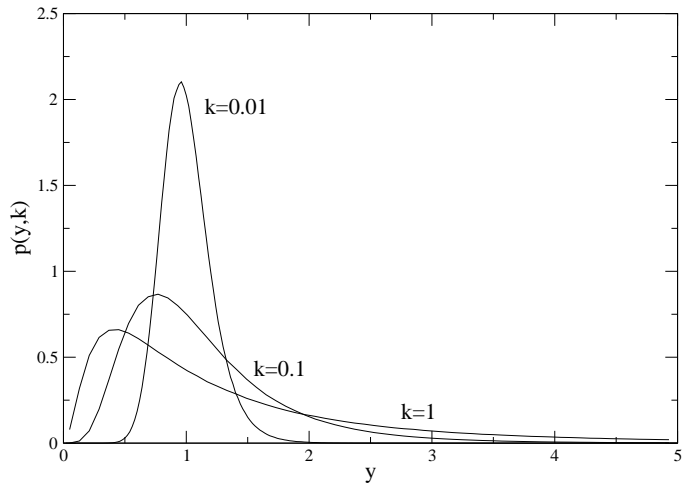}
\caption{Probability density for $\alpha$ for $k=1,\ 0.1, \  0.01$ for $\epsilon(z)=\exp(\phi(z))$ where
$\phi$ is the Gaussian process with correlation function given by Eq. (\ref{gaspr}).
As for the OU process the distribution $p(y,k)$ becomes sharply peaked around $y=1$ which is 
as predicted for small $k$, and $\epsilon^*$ is given by Eq. (\ref{eqstar1})  }
\label{gasmultik}
\end{figure}
We recall that for small $k$ the values of $\alpha _1$ and $\alpha_2$ are 
predicted to be self averaging and thus their distributions should be strongly 
peaked as $k\to 0$.

\begin{figure}
\epsfxsize=0.5\hsize \epsfbox{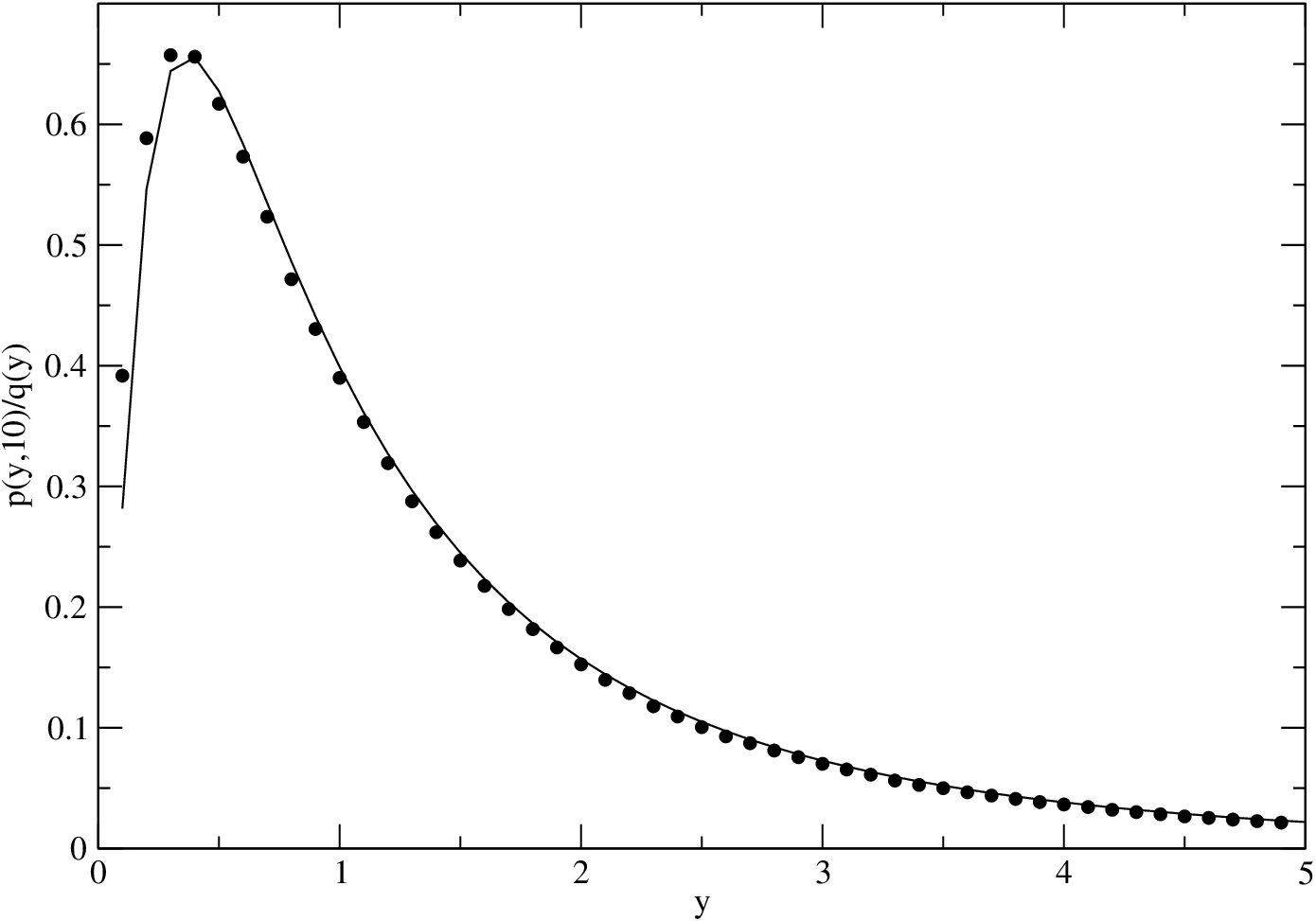}
\caption{The filled circles are the probability density for $\alpha$ for $k=10$ for $\epsilon(z)=
\exp(\phi(z))$ here $\phi$ is the  OU Gaussian process with correlation function given by 
Eq. (\ref{oupr}) with $\omega=1$. The solid line is the probability density function $q(y)$ given in
Eq. (\ref{qy}) for the same distribution of $\epsilon$. The two distributions are already very close 
for $k=10$ in agreement with the prediction of Eq. (\ref{largek}).}
\label{ouk10}
\end{figure}

In Fig (\ref{oumultik}) we have simulated the system when $\epsilon$ is given 
by Eq. (\ref{epsg}) and $\phi$ is an OU Gaussian process (with $\lambda=1$ and 
$\epsilon_g=1$ in Eq. (\ref{epsg}) and $\omega=1$ in Eq. (\ref{oupr})) for values of $k=1$ to $0.01$. 
We see that as predicted the distribution of $\alpha$ becomes increasingly 
peaked about $\epsilon^*=1$ (as given by Eq. (\ref{eqstar1})) as the value of $k$ 
is decreased.  For large $k$ the prediction Eq. (\ref{largek}) can be verified. 
Shown in Fig. (\ref{ouk10}) is the distribution for the same distribution of 
$\epsilon(z)$ but for $k=10$, we see that, as predicted, the probability density 
function for $\alpha$ is already very close to that of $\epsilon$ which is
given by
\begin{equation}
q(y) = \frac{1}{\sqrt{2\pi}y}\exp\left(-\frac{1}{2}\ln^2(y)\right). \label{qy}
\end{equation}
Note that $q(y)$ is independent of the correlation function of the Gaussian field
in the log-normal distribution. 

One can also consider the case of non-Markovian log-normal dielectric 
constants. For instance one can take the field $\phi$ in Eq. (\ref{epsg}) to 
have correlation function

\begin{equation}
\langle \phi(z)\phi(z')\rangle = \exp(-{\omega^2 (z-z')^2/ 2})\;.
\label{gaspr}
\end{equation}

\begin{figure}
\epsfxsize=0.5\hsize \epsfbox{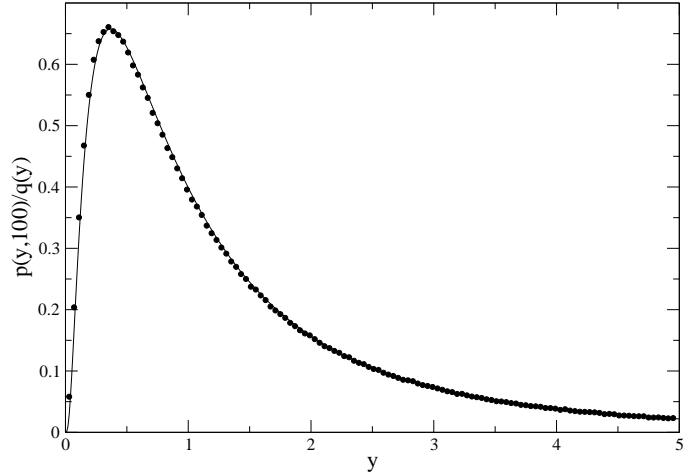}
\caption{The filled circles are the probability density for $\alpha$ for $k=100$ for $\epsilon(z)=
\exp(\phi(z))$ here $\phi$ is the Gaussian process with correlation function given by 
Eq. (\ref{gaspr}) with $\omega=1$. The solid line is the probability density function $q(y)$ given in
Eq. (\ref{qy}) for the same distribution of $\epsilon$. The two distributions are already very close 
for $k=10$ in agreement with the prediction of Eq. (\ref{largek}).}
\label{gas100}
\end{figure}

In what follows we fix $\lambda =1$ and $\omega=1$. For the small values of $k$ 
shown in Fig. (\ref{gasmultik}) we see again that as $k\to 0$ that the 
distribution becomes peaked about $\epsilon^*=1$ as predicted. In Fig 
(\ref{gas100}) we show the comparison between the distribution of $\alpha$ and 
$\epsilon$, again at this value of $k$ the agreement between the two 
distributions is already excellent.

\begin{figure}[t]
\epsfxsize=0.5\hsize \epsfbox{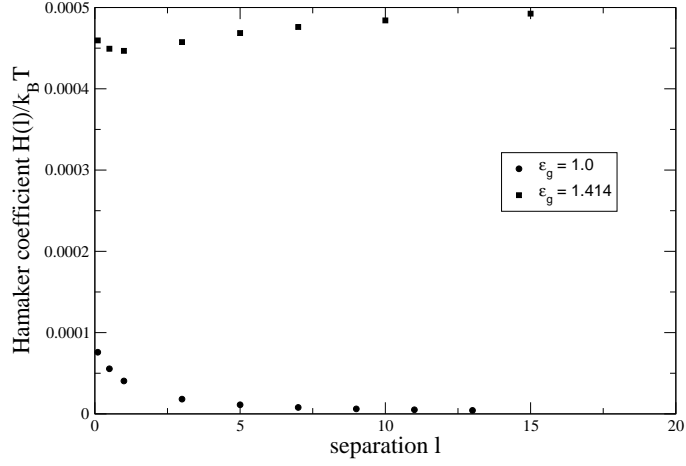}
\caption{The effective Hamaker coefficient $H(l)$ defined in Eq. (\ref{eff_hamaker}) for the distribution
defined by Eqns. (\ref{epsg}) and (\ref{gaspr}) for $\epsilon_g=1.0,~1.414$. In both cases the curves
are asymptotic to the Hamaker constant for a homogeneous system with dielectric constant 
$\epsilon^* = \sqrt{\langle \epsilon \rangle/\langle 1/\epsilon \rangle}$. Notably, $\epsilon^*=1$
for $\epsilon_g=1.0$ and we see that $H(l)$ is asymptotic to zero from above showing that the 
force is attractive for all $l$ even in this case. As is also seen, the curves are asymptotic either 
above or below depending on the value of $\epsilon_g$ and is not necessarily monotonic.  
}
\label{hamaker_0}
\end{figure}

\begin{figure}
\vspace{.75cm}
\epsfxsize=0.5\hsize \epsfbox{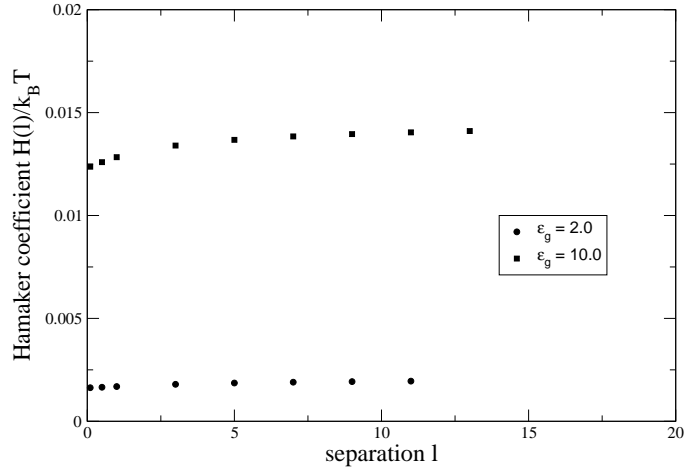}
\caption{The effective Hamaker coefficient $\langle H(l)\rangle $ defined in Eq. (\ref{eff_hamaker}) for the distribution defined by Eqns. (\ref{epsg}) and (\ref{gaspr}) for $\epsilon_g=2.0,~$10.0. In both cases the curves are asymptotic to the Hamaker constant for a homogeneous system with dielectric constant
$\epsilon^* = \sqrt{\langle \epsilon \rangle/\langle 1/\epsilon \rangle}$.
}
\label{hamaker_1}
\end{figure}

We can also compute the average effective Hamaker coefficient as a function of separation $l$ using 
Eq. (\ref{free1}) to define
\begin{equation}
\langle H(l)\rangle  = \frac{k_BT}{16\pi}\int u\,du \int dy_1dy_2\, p_1(u/2l,y_1)p_2(u/2l,y_2)\ln(1-\Delta(y_1)\Delta(y_2)\exp(-u))\;,
\end{equation}
with $\Delta(y) = (y-\epsilon_0)/(y+\epsilon_0)$. For $\epsilon(z)$ distributed according to the 
log-normal distribution in Eq. (\ref{epsg}) and using the scaling result in Eq. (\ref{pscale}) we can
write
\begin{equation}
\langle H(l;\epsilon_g) \rangle =
\frac{k_BT}{16\pi}\int u\,du \int dy_1dy_2\, p_1(u/2l,y_1)p_2(u/2l,y_2)
\ln(1-\Delta(\epsilon_g y_1)\Delta(\epsilon_g y_2)\exp(-u))
\label{eff_hamaker}
\end{equation}
where the $p_i(k,y)$ are computed for $\epsilon_g=1$. We show $H(l;\epsilon_g)$ versus $l$ in 
Figs. (\ref{hamaker_0}) and (\ref{hamaker_1}) for various values of $\epsilon_g$.  

\begin{figure}
\vspace{.75cm}
\epsfxsize=0.5\hsize \epsfbox{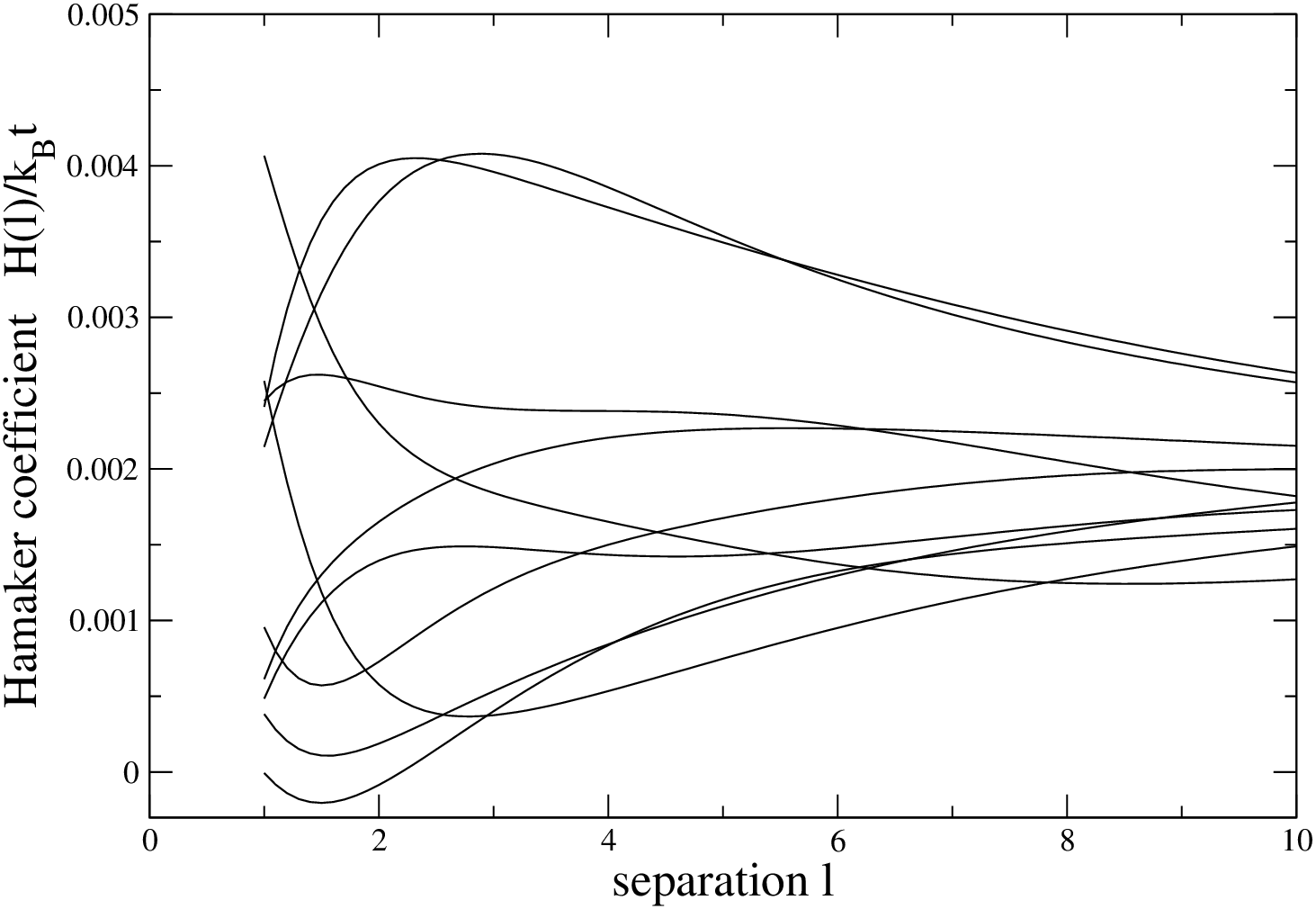}
\caption{Examples of the Hamaker coefficient from the ensemble generated by the distribution for
$\epsilon(z)$ for $\epsilon_g=2.0$ plotted versus separation $l$. The average curve of the ensemble is
shown in Fig. (\ref{hamaker_1}). Note that for small $l$ the curvature of the ensemble curves are not 
of definite sign}
\label{hexamples_2}
\end{figure}

\begin{figure}
\vspace{.75cm}
\epsfxsize=0.5\hsize \epsfbox{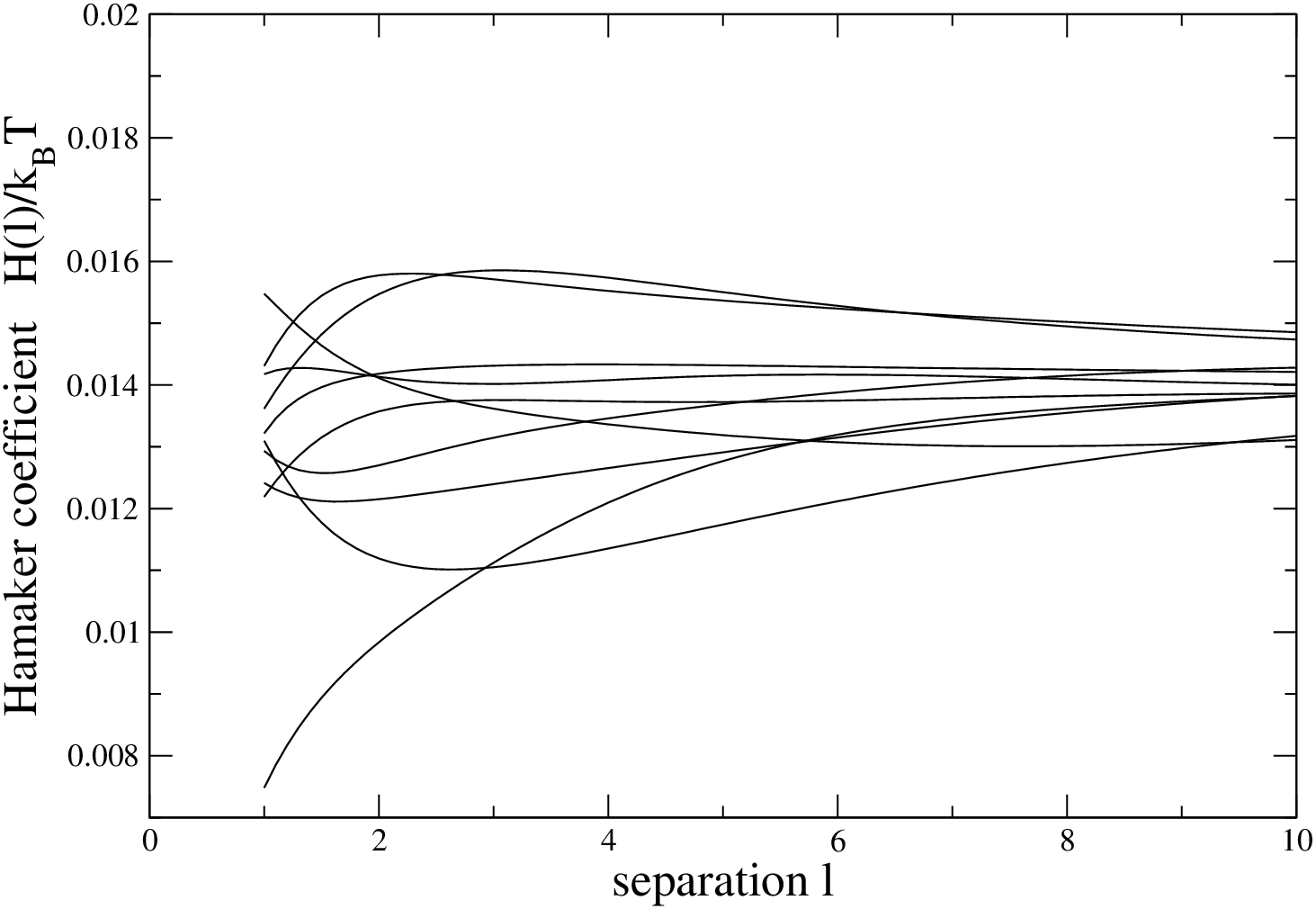}
\caption{Examples of the Hamaker coefficient from the ensemble generated by the distribution for
$\epsilon(z)$ for $\epsilon_g=10.0$ plotted versus separation $l$. The average curve of the ensemble is
shown in Fig. (\ref{hamaker_1}). Note that the curvatures of the ensemble curves are not of definite sign.}
\label{hexamples_10}
\end{figure}

As can be seen $H(l)$ is asymptotic to the  Hamaker constant for a homogeneous system with
dielectric constant $\epsilon^* = \sqrt{\langle \epsilon \rangle/\langle 1/\epsilon \rangle}$,
and that for $\epsilon_g=1.0$ that $H(l)$ is asymptotic to zero from above showing that the
force is attractive for all $l$. This need not have been the case since there will be contributions
from configurations were the dielectric constants $\epsilon_1,\epsilon_2$ in the slabs satisfy
$\epsilon_1 > 1 > \epsilon_2$ or vice-versa; such configurations contribute a repulsive
contribution to the force. As is also observed, the curves are either asymptotic from above or below
and are not necessarily monotonic but can have an initial decrease before rising to the
asymptotic value. These are new and significant features due to the random nature of the system.

In Figs (\ref{hexamples_2}) and (\ref{hexamples_10}) we show a sample of the Hamaker coefficients for particular
realizations of $\epsilon(z)$ plotted against separation $l$ and for $\epsilon_g=2.0,10.0$, respectively. 
The averages over the ensembles from which these curves are taken are given in 
Fig. (\ref{hamaker_1}). It is important
to note that the curvatures of the ensemble curves are not of definite sign and that for small $l$ there are both curves  that decrease and curves that increase with $l$. These properties are reflected in the shape of the  curve for the ensemble average.

\section{Conclusions}

We  have obtained the limiting behavior of the thermal Casimir effect in the 
limit of large separation between the slabs, where the free energy is given by 
self-averaging and thus the distributions of $\alpha(k,z)$ are strongly peaked.
We have shown  that the interaction between  two homogeneous media is stronger than that between  the two fluctuating media if $\langle1 /\epsilon\rangle^{-1} > \epsilon_0$. However it is always weaker  if $\langle \epsilon\rangle <\epsilon_0$.  We thus see that, depending on the details of the distribution of the fluctuating dielectric response in the two slabs and the dielectric response of the medium in-between, the effective  interaction at large inter-slab separations can be stronger or weaker than that for a uniform medium with a dielectric function equal to the average value of that of the media in which it fluctuates. In the limit of small separation, where the interaction free energy is a random variable that has to be averaged over the probability density function of the dielectric functions in the media composing the two interacting slabs. 

The intermediate length scales were analyzed {\em via} perturbation theory and models of disorder that can be treated numerically. All numerical simulations completely corroborate the analytical results for self-averaging at large separations 

The non-linear formulation of the problem presented here should be equally useful to treat the case of deterministically varying dielectric functions and could open up a useful computational framework for designing materials where the effective interaction can be tuned, to induce attractive or repulsive forces depending on the separation \cite{apps}. The formulation also means that if one knows the 
coefficients $a_i(k)$, $a_f(k)$ and $b_j(k)$ for any set of slab media, then one can immediately compute
the effective interaction between them at any distance. This is a rather surprising result as if one 
wanted to compute the effective interaction between two media (1) and (2) using the pair wise approximation, $1/r^6$ for the van der Waals interaction,  it is clear that medium
(1) needs to know what is in medium (2) at each point in order to compute the force. The decomposition
in terms of Fourier modes however means that the  interaction between the two media is
effectively factorized.   A number of further applications of our method would be  to examine the role of disorder for the non-zero frequencies corresponding to  quantum fluctuations and also for different geometries, such as cylindrical and spherical, when the dielectric function varies only in the radial direction. An interesting open problem concerns what happens when the dielectric function varies
in all directions, it would be interesting to know if one can prove in this case whether the long-distance
interaction between two  slabs is also given by the same effective medium expression as derived here.

\begin{acknowledgments} 
This research was supported in part by the National 
Science Foundation under Grant No. PHY05-51164 (while at the KITP program {\em 
The theory and practice of fluctuation induced interactions}, UCSB, 2008). D.S.D 
acknowledges support from the Institut Universtaire de France. R.P. would like 
to acknowledge the financial support by the Agency for Research and Development 
of Slovenia, Grants No. P1-0055C, No. Z1-7171, No. L2- 7080. This study was 
supported in part by the Intramural Research Program of the NIH, Eunice Kennedy Shriver National  Institute of Child Health and Human Development.

\end{acknowledgments}

\pagestyle{plain}

\end{document}